\documentclass[onecolumn]{aa}
\usepackage{psfig}
\begin{document}
\headnote{Research Note}
   \title{Early ComeOn+ Adaptive Optics Observation of GQ Lup and its Substellar Companion
\thanks{Based on observations collected at the European Southern Observatory, La Silla, Chile (ESO Prog ID 53.7-0122).
}
}

   \author{Markus Janson\inst{1} \and
          Wolfgang Brandner\inst{1} \and
          Thomas Henning\inst{1} \and
          Hans Zinnecker\inst{2}
          }

   \offprints{Markus Janson}

   \institute{Max-Planck-Institut f\"ur Astronomie, K\"onigstuhl 17,
              D-69117 Heidelberg, Germany\\
              \email{janson@mpia.de, brandner@mpia.de, henning@mpia.de}
        \and
            Astrophysikalisches Institut Potsdam, An der Sternwarte 16, D-14482 Potsdam, Germany\\
        \email{hzinnecker@aip.de}
             }

   \date{Received ---; accepted ---}

   \abstract{
An analysis of adaptive optics K-band imaging data of GQ Lup acquired in 1994 by the first generation adaptive optics system ComeOn+ at the ESO 3.6m optical telescope in La Silla is presented. The data reveal a likely candidate for the low-mass companion recently reported in the literature. An a posteriori detection in the 11 year old data would provide a useful astrometric data point for the very long period ($\sim 1000$ yr) orbit of the GQ Lup system. However, the data is severely contaminated by speckle noise at the given projected separation, which decreases the confidence of the detection. Still, from the data we can conclude that GQ Lup B is not an unrelated background source, but instead a physical companion to GQ Lup A. We present here the reduction and analysis of the ComeOn+ images, as well as the results. We also discuss the nature of the companion based on data and models available in the scientific literature and examine claims made regarding the classification of the object as a planet.
   }

\authorrunning{Janson et al.}
\titlerunning{Early AO observations of GQ Lup}

   \maketitle

%

\section{Introduction}

Due to advances mainly in the field of adaptive optics (AO), it has been
possible over the last few years to detect substellar-mass companions to stars
at increasingly small separations, and at higher contrast. We are presently at
a point where planetary mass companions could be detected in young systems, if
they are at a sufficiently large projected separation. However, due to
uncertainties in theoretical models and age estimates, large ambiguities can
easily occur in the estimated mass and thus in the classification of an
object. 

The first confirmed detection of a planetary mass object orbiting the brown
dwarf 2MASSWJ1207334-393254 was recently reported by Chauvin et al. (2004,
2005). Neuh\"auser et al. (2005a, henceforth referred to as N1) reported on
the confirmed detection of a companion to the young K-star GQ Lup. The mass
estimate given for this companion was 1 to 40 $M_{\rm jup}$. Since planets are
usually defined as having masses lower than the deuterium-burning limit of
about 13 $M_{\rm jup}$ (sub-stellar objects above this mass being brown
dwarfs), the classification of the object was left an open issue by the
authors. In a subsequent publication (see Mugrauer \& Neuh\"auser, 2005) and
in conference proceedings (see Neuh\"auser et al. 2005b, henceforth N2), they
argued for the interpretation of the object as a planet rather than a brown
dwarf. 

Since 1994, we have carried out high-angular resolution, ground-based adaptive
optics (AO) and HST observations of young stars with the aim to search for
close stellar and substellar companions (e.g.\ Brandner et al.\ 1995, 1997,
2000 and Masciadri et al., 2005). As part of this campaign, GQ Lup was
observed in 1994 using the ComeOn+ adaptive optics system. Given the
brightness and large separation from its stellar companion, GQ Lup B should in
principle be detectable a posteriori upon re-examination of the ComeOn+
data. Here we present the result of such an examination. We also examine the
nature of the source using known measurements and data from literature and
discuss the results. 

\section{Observations and data analysis}

\begin{table*}[htb]
\caption[]{Observing log of high-angular resolution imaging observations of GQ Lup}
         \label{tab1}
\begin{tabular}{rclcllll}
            Date & MJD & Telescope/Instrument  & Filter     &  sep & PA & $\Delta$K & Reference  \\
                 &     &  &                    & [mas] & [$\deg$] & [mag] & \\
            \noalign{\smallskip}
            \hline
            \noalign{\smallskip}
2.\ April 1994   & 49445 &  ESO 3.6m/ComeOn+ & K & 713.8$\pm$35.5 $^a$&  $275.5\pm$1.1 $^a$& $6.24 \pm 0.13$ & this paper \\
10.\ April 1999  & 51279 & HST/WFPC2 & F606W/F814W & $739 \pm 11$ & $275.62 \pm 0.86$ & not given &  N1\\
17.\ July 2002  & 52473 & Subaru/CIAO & K \& L' & $736.5 \pm 5.7$ & not given & not given & N1\\
25.\ June 2004   & 53182 & VLT/NACO  & K & 732.5$\pm$3.4& $275.45 \pm 0.30$ & $6.1 \pm 0.1$ & N1\\
25.\ Aug. 2004   & 53243 & VLT/NACO  & K & 731.4$\pm$4.2& not given & not given & N1\\
14.\ Sept. 2004 & 53263 & VLT/NACO  & K & 735.8$\pm$3.7& not given & not given & N1\\
            \noalign{\smallskip}
            \hline
\end{tabular}
\begin{list}{}{}
\item[$^{\mathrm{a}}$] Error bars refer to statistical error. See the text for discussion on possible biases.
\end{list}
\end{table*}

   \begin{figure}[htb]
   \centering
   \psfig{figure=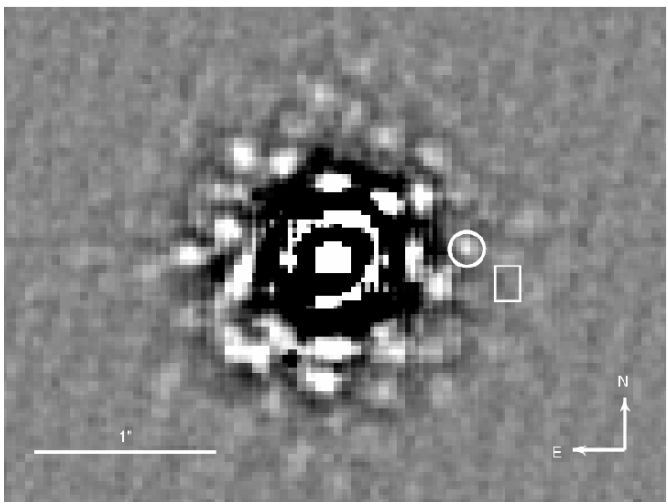,angle=0,width=8.8cm}
\caption{K-band image of GQ Lup after low-frequency filtering has been applied. The circle marks the location of a possible companion candidate. The square marks where the object would be expected if it had been a background source. The image was taken with the ComeOn+ instrument.}
\label{fig1}
    \end{figure}

The observations of GQ Lup were obtained on April 2, 1994 with the adaptive optics (AO) system ComeOn+ and the near infrared camera Sharp2 at the ESO 3.6m telescope in La Silla, Chile. The image scale was 0.05\,arcsec/pixel, and observations were obtained with a K-band filter in order to benefit best from the AO correction. For sky subtraction, GQ Lup was alternatively observed in the lower left, and upper right quadrant of the detector. Individual exposure times were 1\,s, and a total of 80 frames were obtained.

Astrometric calibrations were achieved through observations of the astrometric binary IDS\,17430S6022 (van Dessel \& Sinachopoulos 1993). The observations yield an image scale of 50.15$\pm$0.10 mas/pixel, and that the detector y-axis was aligned with the North direction to within  $\pm$0.20 deg.

The first step in the data reduction consisted of using eclipse tasks (Devillard, 2001) to slice the data cubes into individual 2-dimensional fits frames, and to apply background subtraction and flat field corrections.

While the AO correction in general was good and stable, not all frames have the same Strehl ratio (SR). The average SR was 0.21, while the single best SR achieved was 0.33.  In order to optimize the signal to noise for faint sources located in the point spread function (PSF) wings of GQ Lup, we applied an image selection, selecting a subset of 20 frames with good SR. More details on the rational behind image selection of ComeOn+ and other AO data can be found, e.g., in Tessier et al.\ (1994) and Brandner et al.\ (1995).

The selected frames were then registered (in order to remove residual tip-tilt errors) and combined with the drizzle task in IRAF, selecting a 2-times oversampling of the data. In order to further facilitate the detection of faint companions, a high-pass filter was applied to the data, effectively removing the smooth, sloping background in the PSF wings, and revealing potential companions. Fig. \ref{fig1} shows the resulting image. Using a priori knowledge of roughly where GQ Lup B should be located with respect to GQ Lup A from N1, we found one unique candidate for the object among the many object-like features (due to speckle noise) in the data. 

\subsection{Analysis of the suspected companion}

We calculated separation, position angle and magnitude difference between the
companion and the primary by fitting Gaussians and background to both objects
using the imexamine task in IRAF. The properties of the primary were obviously
measured in the non-filtered counterpart to the image shown as they are
affected by the high-pass filtering. The flux of the secondary was also
calculated in the non-filtered frame, as it will also be affected to a limited
extent by the filtering, which would influence the relative photometry. 

Error estimations based on results from individual frames are particularly difficult when a strong impact of speckle noise is present. This is in part due to the fact that a noise speckle, in difference from e.g. photon noise, forms a coherent structure in space during its lifetime (for an extensive analysis of speckle noise see Racine et al., 1999). This means e.g. that if a noise speckle ends up close to the source in a frame or series of frames, the centroiding procedure will be skewed due to the slope introduced from the speckle in a way that can not be seen as random noise with a Gaussian distribution. In addition, in a case where e.g. the source coincides with a local minimum in the speckle pattern, the centroiding will fail altogether. This means that the frame in question can not be used in the calculations which in turn means that the estimations will be biased towards certain realizations of the speckle pattern.

In order to avoid bias to the greatest possible extent, instead of calculating the errors based on centroiding for each frame, we divide the frames in two groups of the first and last ten frames in the set, respectively. From the difference between the averaged frames of the two groups we estimate the error. The frames for this estimation were high-pass filtered. Obviously, an error estimate based on statistics from two data points is largely statistically uncertain in itself. This is still preferable to the biased estimates from centroiding in each frame (which for this case give lower statistical errors).

Using this method, we get a separation of $713.8 \pm 35.5$ mas, a position angle of $275.5 \pm 1.1$ deg and a magnitude difference of $\Delta$K$= 6.24 \pm 0.13$ mag. In N1, the magnitude difference is $\Delta$K$= 6.1 \pm 0.1$ mag. The values agree within the statistical errors. It should be noted that $\Delta$K can vary systematically, because GQ Lup A is a variable source (see section \ref{sect3}). 

The statistical error is not the greatest danger in interpreting the data. Since some speckle features are rather persistent in the data (present during a large range of frames), they form coherent structures also in the averaged data. Misclassification of any given remaining speckle as an object is as such not a problem, since we have a priori information about the vicinity of the companion which only leaves one reasonable candidate in this case. Rather, the problem is that interference between the source and the remaining speckle pattern could in principle cause severely biased measurements. 

To test the extent of this effect we introduced three artificial structures (Gaussians) in the data with the same expected flux and separation from the star as the real source. For two of these cases, the fake source happened to be positioned near local maxima in the speckle pattern. This led to a brightness which was 2-3 times too high. For the third case, the measured brightness matched the expectation from the known properties of the fake source. All sources had large variations in the measured flux between different frames, showing that they continuously interfere with a dynamical speckle structure.

For the real source, positive interference with a speckle is obviously not a problem, because no such overly bright feature is present where the source should be. However, it can not be entirely excluded that the real source e.g. happens to coincide with a local minimum in the speckle pattern, such that it is hidden in the final data. In this scenario, we would be measuring on a neighbouring local maximum of the speckle pattern, which would essentially mean a non-detection of the source itself. For this reason, rather than claiming a detection of the source in the ComeOn+ data, we postulate that the structure observed is the most probable candidate for GQ Lup B.

\subsection{Examination of the background hypothesis}

Another interesting investigation that can be performed in the ComeOn+ data is to examine the point (relative to GQ Lup A) where GQ Lup B would be expected to appear if it had been a background source. A null detection at this position would further reinforce the (already very solid) conclusion in N1 that GQ Lup B shares a common proper motion with GQ Lup A, and thus also strengthen our a priori assumption of where GQ Lup B should be positioned. Such an examination can be done with much higher confidence than in the common proper motion case, because the hypothetical non-common proper motion case leads to a significantly greater separation at the epoch of the ComeOn+ data aquisition (see Fig. \ref{fig1} and Fig. \ref{fig2}). At the separation of a non-common proper motion case, the impact of speckle noise, and thus potential bias, is much smaller. In addition, the position is further out in the PSF wings of the primary, making background estimation and subtraction less cumbersome.

For this purpose we use the weighted mean of several measured proper motions given in Mugrauer \& Neuh\"auser (2005). This gives $\mu_{\alpha} \cos(\delta) = -19.15 \pm 1.67$ and $\mu_{\delta} = -21.06 \pm 1.69$. It is clear that there are discrepancies between many of the measurements beyond their statistical error bars. A null detection in our data is however quite insensitive to the detailed accuracy of these measurements, because the remaining speckles at the greater separation of a background equivalent to GQ Lup B are too dim (by a factor of at least 2) to constitute false positives.

A search in the area where a non-common proper motion equivalent to GQ Lup B would be expected indeed yields no realistic candidates in the final averaged frame. In order to further test whether this is a strong conclusion, we placed another artificial source in the data with the expected properties of GQ Lup B, at the same separation as a hypothetical background object. The measured flux of the artificial source was in excellent agreement with expectations and with much smaller variations between frames than in the less separated cases. This leads us to conclude that GQ Lup B being a background object can be safely excluded from the ComeOn+ data.

\section{Discussion}

\subsection{The data}

\begin{figure}[htb]
   \centering
   \psfig{figure=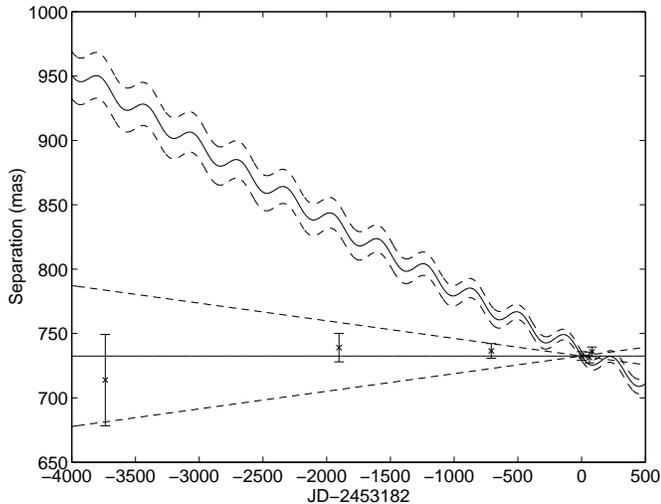,angle=0,width=8.8cm}
\caption{A recreated figure from N1 with inclusion of data from the most probable candidate in the ComeOn+ image. Straight lines indicate expected parameter space for a proper motion companion to GQ Lup A assuming orbital motion of less than $\pm$5 mas/yr (dashed lines are the upper and lower limits, full line is the mean). Wobbly lines indicate the expectation for a background source. The parallax is assumed to be 7.1 mas.}
\label{fig2}
    \end{figure}

\begin{figure}[htb]
   \centering
   \psfig{figure=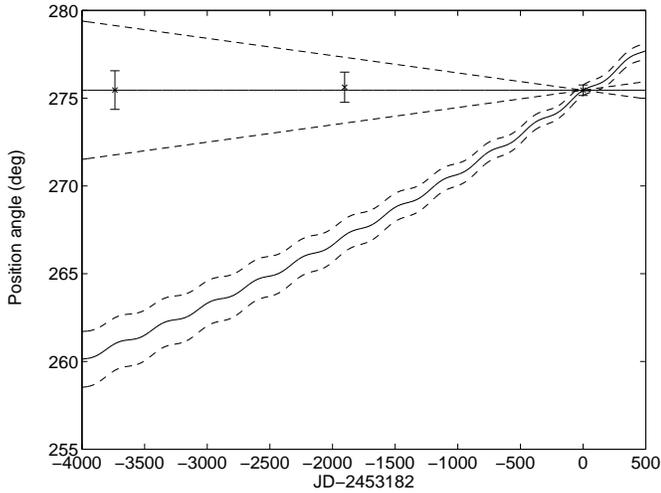,angle=0,width=8.8cm}
\caption{Same as Fig. \ref{fig2} but for position angle instead of separation. The dashed straight lines indicate the expected movement of a companion on a circular, face on orbit of 1000 years. The considerably smaller movement implied by the data may indicate e.g. an inclined orbit, or an elliptical orbit near apastron.}
\label{fig3}
    \end{figure}

It is important to stress the fact that a detection of GQ Lup B in the ComeOn+ data can only be made a posteriori. This is due both to that there are speckle features in the data that mimic companion signatures, as well as that the speckle pattern varies at and around the position of the companion, making its features uncertain. These uncertainties are partly reflected in the rather large error bars of the measurements presented. It is however theoretically possible that bias from speckle influence gives a systematic error beyond the error bars. As mentioned above, the positions of the primary and companion in our data give a separation of $713.8 \pm 35.5$ mas and a position angle of $275.5 \pm 1.1$ deg. The separation is included in a reconstructed diagram from N1 in Fig. \ref{fig2}. A similar diagram has been constructed for the position angle, shown in Fig. \ref{fig3}. 

As previously mentioned, we can exclude that the source is a background object. While it is in principle possible that the sources may share a common proper motion without being physically bound (as independent members of the Lupus association), it is shown in Mugrauer \& Neuh\"auser (2005) that for the case of GQ Lup, this possibility can be rejected with high confidence. Thus it can be safely assumed that GQ Lup A and B are indeed physically bound. 

Due to the large uncertainties in the data and short observational baseline relative to the orbital period of the system, not much can be said about the specific properties of the orbit. However, it can be seen already from the 1999 epoch data and onwards that the position angle changes considerably slower than what would be expected if the orbit was circular and face-on. This implies that the orbit is either elliptical and currently near apastron, or has a significant inclination, or both.

\subsection{The GQ Lup system}

In N1, it is concluded that both planetary and brown dwarf masses are included within the error bars of the mass estimation, and that a classification of the object based on current knowledge is thus not possible. In N2 it is instead concluded, based mainly on a different emphasis on which models to use, that the object is "almost certainly a planet". Using measurements and models from scientific literature, we have investigated the nature of the source in order to add to the discussion on whether or not the companion is a planet.

\subsubsection{Distance}

A first important discussion concerns the distance to GQ Lup. The star has no parallax measured, so any distance estimation has to be made through its highly probable association with the Lupus I molecular cloud. Distance estimates to the Lupus clouds have varied quite a lot historically, but more recently applied methods such as ubvy$\beta$ photometry (see Franco, 1990) and projection of the cloud complex on its background (see Murphy et al., 1986) give a fairly consistent result of $150 \pm 20$ pc. Based on extinction jumps among sources in the direction of the Lupus clouds with known distances, Knude \& Hog (1998) estimate a distance to the Lupus clouds of about 100 pc. However, due to the inconsistency with other results, and since the effect could either way not be shown for the Lupus I cloud, we disregard this result and use $150 \pm 20$ pc as mentioned above. This range is a bit more restrictive than the $140 \pm 50$ pc distance estimate used in N1. Eventually, a parallax measurement will have to be performed in order to get an entirely reliable distance to GQ Lup.

\subsubsection{Brightness and extinction}

\label{sect3}

GQ Lup is a highly variable source. Since the brightness varies, it is difficult to find the "true" brightness of the system, i.e. the brightness that corresponds to the actual radius of the star. This is the brightness which is relevant when placing the star in an HR diagram to deduce its age and mass from modelled evolutionary tracks. Herbst et al. (1999) mention that GQ Lup varies in visual brightness by more than two magnitudes. This is based on measurements (available from ftp://www.astro.wesleyan.edu/pub/TTauri/WTTS/) over three different epochs, giving $V_{\rm min} = 11.33$ mag and $V_{\rm max} = 13.36$ mag. However, we disregard the first epoch since those measurements give no colour index in the ranges that interest us (this has a minimal effect on the brightness estimation). We also disregard the third epoch, where all values with $V \ge 12$ mag are present. This is because those values represent a minority of the data, and values from other sources at other epochs (e.g. Bastian \& Mundt, 1979 and Mundt \& Bastian, 1980) are all within $V = 11$ to 12 mag, implying that the epoch 3 data represents an extreme rather than the normal behaviour of the source. We will however mention in section \ref{sect2} what inclusion of these values would result in.

For T Tauri stars, it is risky to calculate the interstellar extinction based on the colour excess in $B - V$. This is because accretion of material from the disk onto the star causes a UV excess which can have a significant effect on the B-band. Instead, we use colours $V - R$ and $R - I$ (given along with the brightness at the mentioned website) independently to check whether the results are consistent. We find that $A_{\rm R} = -0.062 \pm 0.19$ mag and $A_{\rm I} = -0.031 \pm 0.19$ mag. Since a negative extinction value is obviously impossible, we set the extinction to zero with an upper limit of $A_{\rm R, max} = 0.13$ and $A_{\rm I, max} = 0.16$. Is zero extinction a reasonable result? For GQ Lup, the answer is yes. Over a distance of only about 150 pc, interstellar extinction has a low impact if no dense regions occur between the source and observer. The Lupus I molecular cloud of which GQ Lup is a member is not expected to contribute to the extinction either, for two reasons: Firstly, faint sources in a molecular cloud are more easily detected if they are on the near side of the cluster than on the far side, and thus a certain detected T Tauri star is obviously more likely be on the near side. Secondly, and more importantly, from CO maps of the Lupus clouds (see Tachihara et al., 1996), it is clear that GQ Lup is at the very edge of the Lupus I cloud, and will therefore not experience much extinction from the cloud regardless of depth. In addition, GQ Lup is a weak-line T Tauri star with a $W_{\rm H \alpha}$ of only 2.8 \AA . Thus, the circumstellar disk will also not contribute largely to the extinction.

\subsubsection{Age and mass estimation}

\label{sect2}

Using these values for distance, apparent magnitude and extinction, we derive absolute magnitudes of $M_{\rm R} = 4.92^{+0.40}_{-0.73}$ mag and $M_{\rm I} = 4.16^{+0.60}_{-0.67}$ mag. Using known relations between luminosities and magnitudes in the respective photometric bands for a K7 star (equivalent to applying a bolometric correction), we get nicely consistent values of $L(R) = 0.87^{+0.83}_{-0.37} L_{\rm sun}$ and $L(I) = 0.86^{+0.74}_{-0.36} L_{\rm sun}$. We henceforth use the slightly less restrictive luminosity measure derived from the R-band photometry.

\begin{table}[htb]
\caption[]{Physical properties of GQ Lup A \& B}
         \label{tab2}
\begin{tabular}{ccc}
            Properties & A & B \\
            \noalign{\smallskip}
            \hline
            \noalign{\smallskip}
Distance (pc) & 130-170 & same \\
$A_{\rm R} (mag)$ & 0-0.13 & same \\
Spectral type & K7Ve & M9-L4 \\
$L_{\rm bol}/L_{\rm sun}$ & 0.50-1.71 & 0.0025-0.01 \\
Age (Myr) $^a$& 0.3-5 & same \\
Mass $^a$& 0.4-0.8 $M_{\rm sun}$ & 3-50 $M_{\rm jup}$ \\
            \noalign{\smallskip}
            \hline
\end{tabular}
\begin{list}{}{}
\item[$^{\mathrm{a}}$] Model-dependent estimations. The lower and upper values do not correspond to the overlapping regime of the models, but to the extremes of the combined ranges.
\end{list}
\end{table}

\begin{figure}[htb]
   \centering
   \psfig{figure=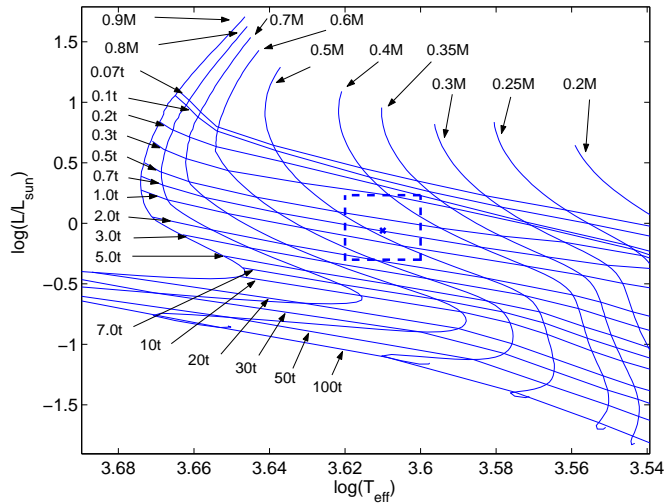,angle=0,width=8.8cm}
\caption{Evolutionary mass tracks from D'Antona et al. (1997). Tracks labelled 'M' are mass tracks in units of $M_{\rm sun}$. Tracks labelled 't' are isochrones in units of Myr. The dashed square indicates the range of luminosities and effective temperatures of GQ Lup A as given in the text.}
\label{fig4}
    \end{figure}

From the spectral type of GQ Lup, we get $\log(T_{\rm eff}) = 3.61 \pm 0.01$. Knowing both the temperature and luminosity, we can place the source in an HR diagram and from theoretical evolutionary tracks for PMS stars get an estimate of the mass and age of GQ Lup. We use two different models for this purpose: Siess et al. (2000) and D'Antona et al. (1997). The Siess et al. (2000) model gives roughly an age range of $7*10^5$ to $5*10^6$ years, whereas the D'Antona et al. (1997) model gives $3*10^5$ to $2*10^6$ years. Given the uncertainty of the models at such low ages, the consistency between the results is reasonable. Compared to the age range suggested in N1 of 0.1 to 2 Myrs, the D'Antona et al. (1997) model gives a consistent age, whereas the age estimated from Siess et al. (2000) is somewhat higher. For mass estimations, the Siess et al. (2000) model gives a range of 0.6 to 0.8 $M_{\rm sun}$ and the D'Antona et al. (1997) model gives 0.4 to 0.6 $M_{\rm sun}$. For both mass and age, we adopt the full range given by the extremes of the two models. It should be pointed out for completeness that if the full range of photometric values measured is taken into account (see the discussion on variablilty in section \ref{sect3}), the upper limit of the age is extended to the point where every age up to the ZAMS is possible. Some D'Antona et al. (1997) evolutionary tracks along with GQ Lup A are shown in Fig. \ref{fig4}.

\subsubsection{The nature of GQ Lup B}

Having calculated some of the fundamental properties of the GQ Lup system using the A component, we can attempt to establish the nature of GQ Lup B/b. Using the photometric data given in N1, and the distance and extinction of this paper, using the same procedure as above, we get a logarithmic luminosity of about $\log(L/L_{\rm sun}) = -2.3 \pm 0.3$. We adopt the same broad temperature range as N1 of 1200 to 2500 K.

Using the luminosity and temperature given, along with the age calculated for the primary (assuming an equal age of the components in the system), we can estimate the mass with theoretical models, though it is highly unclear to which extent present models can be applied to such a low-mass object, and in particular for such a young age range. From Fig. 1 in Baraffe et al. (2003) and the corresponding table at perso.ens-lyon.fr/isabelle.baraffe, using luminosity and age we get masses from $\sim 12$ $M_{\rm jup}$ to $\sim 40$ $M_{\rm jup}$. Using temperature and age we get masses from $\sim 3$ $M_{\rm jup}$ to $\sim 20$ $M_{\rm jup}$. The results leave both a planetary and a brown dwarf interpretation of the object open. We thus conclude that with the data present it is not possible to determine the classification of GQ Lup B, and that improvements must be made in both measurements and models for such a step to be taken. This conclusion is very similar to the one drawn in N1, aside from that they also present results from an extension of a model of Wuchterl et al. (2003) which gives masses which are all well below the planetary limit.

In N2 there is instead a much stronger emphasis on the classification of GQ Lup B/b as a planetary object, and it is eventually claimed that the object is almost certainly a planet. This statement is based mainly on arguments regarding which models are the most appropriate. The Wuchterl et al. (2003) model is preferred to the Burrows et al. (1997) and Baraffe et al. (2002) models, since it takes physical consideration to the initial cloud collapse. However, as is rightly implied in N1, neither of the models have been properly calibrated for such young, low-mass objects. Even though Wuchterl et al. (2003) give a seemingly more complete physical picture of the evolution of PMS stars by including the protostellar collapse, this is certainly not a guarantee that the results yielded will be necessarily better. This must be tested against, e.g., dynamical observations of bodies in the relevant mass and age range. We discuss some further aspects of the issue of evolutionary models for the case of GQ Lup B in section \ref{sect1}.

Other arguments in N2 concern results reached by comparison of the measured spectra of the companion with artificial spectra generated by the GAIA-dusty model, and results presented in Mohanty et al. (2004). The latter results are also based on spectral models. Both these models are extensions of the Allard et al. (2001) model, but neither of the two has been published in a refereed context. The Allard et al. (2001) models had severe problems in some cases for young low-mass objects (see Burgasser et al., 2004). Even though the updated models contain more detailed physics such as updated (but still incomplete) line lists to account for these problems, this fact serves to caution that it is dangerous to apply such models in ranges for which they have not been tested. Again, dynamical observations are needed to calibrate the models.

In N2 a statement is made which can be easily interpreted as that according to both Mohanty et al. (2004) and Close et al. (2005), the models of Baraffe et al. (2002) underestimates the masses of objects with actual masses $> 40 M_{\rm jup}$ but overestimates them for objects $< 30 M_{\rm jup}$. While Mohanty et al. (2004) do indeed state this, it is important to note that Close et al. (2005) make the opposite assumption (based on extrapolation of the observed trend) about masses below their object of study, which is AB Dor (about 90 $M_{\rm jup}$). This is an important distinction, since the mass of AB Dor is determined dynamically. It is found that theoretical mass tracks underestimate the mass of AB Dor by a factor of about 2.5 (but see also e.g. Luhman et al., 2005 and Nielsen et al., 2005 for further discussions). Another survey of dynamical masses versus evolutionary tracks of young, low-mass stars (down to 300 $M_{\rm jup}$) performed by Hillenbrand and White (2004) shows a similar trend, namely that the masses tend to be underestimated by theoretical tracks. Whether this trend can be extrapolated to even lower mass objects such as GQ Lup B/b or whether there is an opposite trend as suggested by Mohanty et al. (2004) is still very much an open issue from a dynamical point of view. Suitable sources for such comparisons, which can be observed both photometrically and dynamically, are still lacking.

Another point needs to be mentioned regarding the presentation of results from Mohanty et al. (2004) in N2. It is claimed that the faintest object from Mohanty et al. (2004) is about 9 $M_{\rm jup}$, and that since GQ Lup B/b is both colder and fainter, it must be lower in mass, giving a mass estimate of $\le 8$ $M_{\rm jup}$. However, it is clear from Mohanty et al. that the errors are within a factor of two. Thus the upper limit is 18 $M_{\rm jup}$, and so the mass estimate of GQ Lup based on this would be $\le 17$ $M_{\rm jup}$, which includes super-planetary masses.

As a final comment on the issue of the physical nature of GQ Lup B, it should be noted that the object does not seem to constitute a new class of its own. There exist several detected objects with observed properties similar to those of GQ Lup B, for instance HR 7329 B, GG Tau B and AB Pic B. If GQ Lup B is to be classified as a planet, then it is possible that several of these objects are planets as well. This abundance of possible borderline cases further underlines the fact that it is important to validify the theoretical models before making a definite statement about the nature of these objects.

\subsubsection{The history of the system}

\label{sect1}

A recent pre-print by Fortney et al. (2005) presents a model for the formation of $2 M_{\rm jup}$ planets based on the core accretion scenario. It is found that with this model, the planets become fainter than what the model of e.g. Baraffe et al. (2003) gives for the same mass at young ages. This means that according to the Fortney et al. (2005) model, the Baraffe et al. (2003) model underestimates the masses of young, low-mass objects, which is opposite to the conclusion of Wuchterl et al. (2003). It is worth noting that for masses below $13 M_{\rm jup}$, N1 use a modified version of Wuchterl et al. (2003) which is based on nucleated instability, since the object can not collapse by itself for such low masses in this model. In other words, the modified model assumes a core accretion formation scenario, just like the model of Fortney et al. (2005). However, Fortney et al. (2005) specifically treats the runaway accretion of gas onto the core in the context of the solar nebula, whereas the modified Wuchterl et al. (2003) model in N1 treats the gas collapse onto the core with the same boundary conditions as for a regular protostellar cloud collapse (or so we interpret the description in N1). Thus possibly, the Fortney et al. (2005) model is more appropriate for objects which form in a circumstellar disk. Of course, the Fortney et al. (2005) and Wuchterl et al. (2003) models both suffer from the same lack of dynamical calibration already mentioned.

This discussion inevitably brings to light the issue of how and where GQ Lup B was formed. Giant planets are highly unlikely to form at a separation of about 100 AU, since the disk is not sufficienly dense so far out. It is not impossible that GQ Lup B formed in the disk of GQ Lup A at a smaller separation and was subsequently thrown outwards through e.g. a planet-planet interaction. On the other hand, theoretical star formation models based on turbulent fragmentation (see e.g. Padoan \& Nordlund, 2002) indicate that objects far below the stellar mass limit can form independently in the same way as ordinary stars. It is hardly relevant to refer to objects formed in such a way as planets, regardless of their final mass. Thus if such a formation scenario is applicable down to a few Jupiter masses (some arguments for this case are presented in Chabrier et al., 2005), this may affect also the appropriate classification of GQ Lup B, given its peculiar orbit. In addition, of course, the assumption of co-evality would not necessarily hold in such a scenario, making the age of GQ Lup B unknown.

There is also a timescale issue. As we have seen, the N1 variant of the Wuchterl et al. (2003) model presumes that GQ Lup B formed by the core accretion-gas capture scenario. However, such a scenario likely takes place over a timescale on the order of at least $10^6$ yrs. It is not obvious whether such a timescale is compatible with the low estimated age of the GQ Lup system. It should be mentioned in this context that if GQ Lup B despite the timescale and orbital problems did form in the disk of GQ Lup A through core accretion, then it is considerably younger than GQ Lup A, rather than co-eval. In such a case, an otherwise accurate model would overestimate the mass of GQ Lup B under the assumption of co-evality.

\section{Conclusions}

The strong impact of speckle noise in our eleven year old data leads to a weak conclusion regarding the detection and properties of GQ Lup B. The only candidate object in the data has reasonable flux and position values. It can however not be entirely excluded that the detection is due to persistent bias in the data.

A stronger conclusion based on the ComeOn+ data that can be made is that GQ Lup B is not a background object since no matching object can be found near the position relative to GQ Lup A where such an object would be expected. The impact of speckle noise is considerably lower at this position, which means that if there was a source there with the brightness of GQ Lup B, we could indeed expect to see it.

Based on huge uncertainties in both currently available data and models, we conclude that it is impossible to classify GQ Lup B exclusively as a planetary mass object or a brown dwarf. Efforts which could eventually help resolve this ambiguity include dynamical measurements of the GQ Lup system, a parallax measurement of the system, an extensive study of the variability of GQ Lup A, and dynamical calibrations of evolutionary and spectral models from other similar objects.

\begin{acknowledgements}

Support was provided by the Deutsches Zentrum f\"ur Luft- und Raumfahrt (DLR), F\"orderkennzeichen 50 OR 0401. 

\end{acknowledgements}

\end{document}